\documentclass[aps,twocolumn,secnumarabic,nobalancelastpage,amsmath,amssymb,longbibliography]{revtex4-1}
\usepackage{amsmath}
\usepackage{graphicx}      
\usepackage{longtable, gensymb, xcolor}     
\usepackage{url}           
\usepackage{bm}            
\usepackage[english]{babel}
\usepackage{float}

\begin{document}
\title{Macroscopic Dark Matter
Constraints from Bolide Camera Networks}
\author         {Jagjit Singh Sidhu, Glenn Starkman}
\affiliation    {Physics Department/CERCA/ISO Case Western Reserve University
Cleveland, Ohio 44106-7079, USA}
\date{\today}

\begin{abstract}
Macroscopic dark matter (macros)
are a broad class
of alternative candidates to particle
dark matter.
These candidates would transfer
energy
primarily through 
elastic scattering, 
and this energy deposition
would produce observable signals
if a macro were to pass
through the atmosphere.
We produce constraints for 
low mass macros from the null
observation of bright meteors formed
by a passing macro, across two
extensive networks
of cameras built
originally to observe meteorites.
The parameter
space that could be probed
with planned upgrades to the existing
array of cameras in one
of these networks still
currently in use, 
the
Desert Fireball Network in Australia,
is estimated.

\end{abstract}
\maketitle
\section{Introduction}
Assuming General Relativity is the correct
theory of gravity on all scales, 
there is considerable evidence for dark matter.
Macroscopic dark matter (macros)
is a broad class of dark-matter candidates, 
with wide still-allowed  ranges
of  masses $M_x$ and cross sections $\sigma_x$,
that represents
an alternative to conventional particle dark matter.

Of particular interest would be macros
of approximately nuclear density
satisfying
\begin{equation}\label{nuclear}
\sigma_x \approx 2\times 10^{-10} 
\left(\frac{M_x}{g}\right)^{\frac{2}{3}}
{cm^2}\,,
\end{equation}
as several models for macros
describe potential candidates
with approximately that density.
The idea that macros
could be formed entirely within
the Standard Model was originally
proposed by Edward Witten \cite{PhysRevD.30.272}
in the context of a first-order
QCD phase transition. 
Lynn, Nelson and Tetradis \cite{LYNN1990186} 
and Lynn again \cite{1005.2124} 
subsequently described a more 
realistic model for Standard-Model macros 
as bound states of nucleons 
with significant strangeness.
Nelson \cite{Nelson:1990iu}
studied the formation of nuggets of strange-baryon matter
during a second QCD phase transition --
from a kaon-condensate phase
to the ordinary phase. 
Others have considered 
non-Standard-Model versions of 
such objects and their formation
\cite{Zhitnitsky2003}.
Although macros satisfying \eqref{nuclear}
are of particular interest, macros over a wide range of 
average densities remain possible candidates to explain
the open problem of the nature of dark matter.

Colleagues  
recently determined which regions of macro
parameter space remain unprobed \cite{jacobs2015macro,jacobs2015resonant}.
A longstanding constraint comes from
examination of a slab of ancient mica
for tracks that would have been
left by the passage  of a macro moving 
at the typical speed of dark matter in the Galaxy.
This was used to rule out macros
of $M_x \leq 55\,$g for a wide range of
cross sections (see
\cite{Price:1988ge} and
\cite{DeRujula:1984axn}).
Various microlensing experiments
have constrained the dark-matter fraction 
for masses $M_x \geq 10^{24}\,$g 
\cite{Alcock2001,Griest2013,astro-ph/0607207,0912.5297}.
Reference \cite{1309.7588} utilized the full Boltzmann formalism to obtain constraints
from macro-photon elastic scattering using the first year release of Planck data.
In particular, a sufficiently high dark-matter photon interaction will
generate distinctive features in the temperature and polarization power
spectra at high $\ell$ values. Constraints were derived by 
comparing the spectra to the latest Planck data, and finding the
best-fit cosmological parameters. 
Prior work had already constrained a similar range of 
parameter space by showing that
the consequence of dark matter interactions with standard model particles is
to dampen the primordial matter fluctuations and essentially
erase all structures below a given scale (see e.g. \cite{Bhm2001}).
More recently, the existence of massive white dwarfs
was used to constrain a significant region
of macro parameter space \cite{1805.07381}. 
The region of parameter space where macros
would have produced a devastating injury similar
to a gunshot wound on the carefully
monitored population of the western world
was also recently constrainted \cite{1907.06674}.

More work has been done recently
to identify additional ways to probe 
macro parameter space.
With colleagues, we have proposed \cite{1808.06978}
using current Fluorescence Detectors
that are designed to study High Energy Cosmic Rays,
such as those of the Pierre Auger Observatory \cite{PAOFD}. 
Separately, we have suggested  \cite{1905.10025}
that, for appropriate $M_x$ and $\sigma_x$,
the passage of a macro through granite 
would form long tracks of melted and re-solidified rock 
that would be distinguishable 
from the surrounding unmelted granite. 
A search for
such tracks in commercially available granite slabs 
is planned.

In this manuscript, we describe a way
to constrain macro parameter space
based on an
idea first put forward by Hills
\cite{Hills1986},
who used the non-observation of fast-moving meteors 
to constrain a wide range of masses 
of atomic-density dark-matter candidates.
We show
that far-denser dark-matter candidates
were also constrained by this ``fireball" 
null result.

The Desert Fireball Network (DFN)
\cite{Howie2017} is a network of cameras in Australia
searching for bright meteors, which are 
known as bolides.
The area covered by
the DFN is already much greater than
the original network used by Hills,
and thus produces more stringent constraints on macros. 
There are plans for significant DFN upgrades, 
including increases in observing area, 
as part of an effort to create a global network. 
These upgrades will yield 
still-more-stringent constraints on macros.
We estimate below the region of parameter space 
that could be probed given some reasonable 
expectations for the eventual size and sensitivity of the
global network of bolide detectors.

\section{Macro Detectability}
{\color{black} In this section, we provide a guide
for how we determine constraints from the lack of 
observation of fast-moving 
bolides in surveys using in the following section.}

A macro 
transiting the atmosphere 
will deposit energy along its essentially straight-line path
through elastic scattering
off  molecules in the air. 
It will do so at a rate
\begin{equation}\label{dedx}
\frac{dE}{dx} = \sigma_x \rho_{atm}(D) v_x^2 \,.
\end{equation}
Here 
$\rho_{atm}(D) \approx 
e^{-D/10{\mathrm{km}}}~{\mathrm{kg~m}}^{-3}$ 
is the density of the atmosphere at altitude $D$, 
accounting for the atmosphere's scale-height
of approximately $10$km, 
$\sigma_x$ is the geometric cross section of the macro, 
$v_x$ is its speed.

This energy deposition along the macro trajectory
will result in some flux of visible light 
$F(\sigma_x, v_x; D)$ 
received at the cameras
that make up a detector array
designed to search for bright meteors. 
These cameras have a minimum flux $F_{thresh}$
for a bolide to be visible.
Macros that result in a flux
greater than the threshold
will produce a detectable signal.
This implies that at each altitude $D$ 
there is a minimum velocity $v_{thresh}(\sigma_x;D)$ 
for a macro of cross-section $\sigma_x$ 
to be detectable. 
In other words, the macro is detectable if 
$v_{x}>v_{thresh}(\sigma_x;D)$ -- 
the smaller $\sigma_x$,
the larger $v_{thresh}$ must be (at any given $D$).

For definiteness, 
we assume macros possess a Maxwellian velocity distribution in a frame co-moving with the Galaxy,
\begin{equation}
	\label{eq:maxwellian}
	f_{MB}(v_x) = 
		\frac{4\pi v_x^2}
		{\left({\pi v_{vir}^2}\right)^{3/2}}~
		e^{-\left(\frac{v_x}{v_{vir}}\right)^2}, 
\end{equation}
where $v_{vir} \approx 250~ \text{km s}^{-1}$\footnote{
	This is the distribution of macro velocities in a non-orbiting frame moving with the Galaxy.
	When considering the velocity of macros impacting the atmosphere, \eqref{eq:maxwellian} is modified by the motion of the Sun and Earth in that frame, and by the Sun's and Earth's gravitational potential. 
	We have taken into account these effects 
	(as explained, for example, in  \cite{Freese2013}),
	except the  negligible effect 
	of Earth's gravitational potential.
	},
but cut off at $v_{x,esc} \sim 550\,$km s$^{-1}$ --
the escape velocity from the Galaxy at the  Sun's position. 
The macro flux 
detectable from altitude $D$
is given by  
the cumulative velocity distribution function -- specifically
the integral of the product of 
the relative velocity
between the macro and the Earth and
the macro velocity-distribution function 
(transformed to the rest frame of the solar system {\color{black}$f_{MBSS}(v_x)$}
from $v_{thresh}(\sigma_x;D)$ 
to $v_{x,esc}$.

The local mass density of dark matter is fixed
by Galactic dynamics.
For simplicity of interpretation, 
we consider all macros to be 
of a single mass and size,
even though a broad mass distribution 
is a reasonable possibility 
in the context of a composite dark-matter candidate.
The macro flux is therefore inversely
proportional to the macro mass, and we must have
\begin{widetext}
\begin{eqnarray}
	\label{fluxcondition}
	\Gamma_{min} 
	 &\leq& 
	\Gamma\left(M_x,\sigma_x\right)\equiv
	A_{det}
		{\color{black}\frac{\rho_{DM}} {M_x} \int_{v_{thresh}(\sigma_x;D)}^{v_{esc}}
		\!\!\!\!\!\!\!\!\!\!\!\!\!\!\!\!\!\!\!\!\!\!
		v_x f_{MBSS}(v_x)  dv_x \,.}
\end{eqnarray}
\end{widetext}
where 
$\rho_{DM}\simeq 5\times10^{-25}\mbox{g}~\mbox{cm}^{-3}$ \cite{Bovy2012},
and $A_{det}$ is the projected area on the sky covered by the 
network of cameras in the bolide survey.
{\color{black}Consequently, the smaller $v_{x,thresh}$,
the lower $\sigma_x$ we can probe at the cost of only
being sensitive to a fraction of the macros
in the Maxwellian distribution, corresponding to a maximum
macro mass.}
{\color{black}The smaller $\sigma_x$ is,
the higher  $v_{x,thresh}$ needs to be to produce
enough photons for an event to be detectable.
That higher $v_{x,thresh}$ means that a smaller fraction of the macros are detectable,
so to have a high enough event rate, 
the overall number density of macros must be higher, i.e. they must be of lower mass.
Higher $\sigma_x$ thus means lower maximum probable mass.
}

Equation (\ref{fluxcondition})
allows us to determine, as a function of $\sigma_x$,
the maximum mass $M_x$ that we can probe,
so long as we know the velocity distribution 
as a function of altitude.
The speed of a macro traveling through
the atmosphere is expected to evolve as
\begin{equation}\label{speedevolution}
v(x) = v_{0} 
	e^{-\langle \rho \Delta\rangle 
	\frac{\sigma_x}{M_x}}\,.
\end{equation}
Here $\langle \rho \Delta\rangle$ is
the integrated column density traversed 
along the trajectory from the point of impact 
to the location x, 
\begin{equation}
	\langle \rho \Delta\rangle \equiv 
	\int_l e^{-{\color{black}D(x)}/10{\mathrm{km}}}~{\mathrm{kg~m}}^{-3} dx\,,
\end{equation}
where $l$ represents the trajectory of the macro
and $10\,$km in the exponent is the atmospheric scale height.

For large enough values of ${\sigma_x}/{M_x}$,
macros will not traverse the Earth and 
only macros at high altitude will produce visible signals.
This could be allowed to determine
the maximum value of
${\sigma_x}/{M_x}$ accessible by the bolide network.
%
%
However, in order to be detected by the camera network,
a macro must remain intact as it passes through the atmosphere. 
Ordinary meteors often break up.

We expect sufficiently dense objects 
to be stable enough
to  survive passage through the upper atmosphere.
The precise threshold density depends on the
microscopic physics of the macro,
however we can get a sense of how dense
by imagining the macro is made of baryons,
and that the logarithm of the binding energy per baryon $E_b$
scales linearly with  the logarithm of the  density 
between atomic density ($\rho_{atomic}\simeq 1g/cm^{-3}$, $E_b\simeq 10$eV)
and nuclear density ($\rho_{nuclear}\simeq 10^{14}g/cm^{-3}$, $E_b\simeq 1$MeV).

{\color{black} 
This yields an expression for the scaling between binding
energy and density
\begin{equation}
E_b \sim 10 eV \left(\frac{\rho_x}{g/cm^{-3}}\right)^{\frac{3}{7}}\,,
\end{equation}
where 
\begin{equation}
\rho_x = \frac{3 M_x \pi^{\frac{1}{2}}}{4 \sigma_x^{\frac{3}{2}}}\,.
\end{equation}
We require the energy transferred be less than
the binding energy per baryon multiplied by the
number of baryons in the macro
\begin{equation}
E_b \frac{M_x}{m_b} \geq \rho \sigma_x v_x^2 L\,.
\end{equation}
Considering just the densest part of the atmosphere, i.e.
that closest to the ground, i.e. $\rho \sim 10^{-4}\,$g cm$^{-3}$
in the scale length closest to the ground, we find the above
expression translates into a bound
\begin{equation}
\frac{\sigma_x}{\mathrm{cm^2}} \lesssim 4 \times 10^{-3} 
\left(\frac{M_x}{\mathrm{g}}\right)^{\frac{20}{23}}\,.
\end{equation}
This bound is likely too stringent in that most macros at these large cross
sections will be detectable at higher altitudes with significantly
lower atmospheric densities. However, we use this as our upper bound
in Figure \ref{fig:bolide}.
}


\section{Detection Thresholds}
A bolide refers to a very bright
meteorite with $M_v \geq -5$. 
The original bolide networks
operating in the 1960s, '70s and '80s 
were reliably capable of detecting objects 
of absolute visual magnitude $M_v \leq -5$
\cite{mccrosky_boeschenstein_1965}.
The absolute magnitude \cite{1959}
of a bolide is related to its luminosity by
\begin{equation}\label{opik}
M_v = 6.8 -2.5 \log_{10} ({\cal L}/{\cal L}_0)
\end{equation}
where ${\cal L}$ is the visible luminosity of a regular
meteor and ${\cal L}_0= 1 W$. Based on the criterion $M_v \leq -5$, 
bolides would have been detectable 
when their luminosity in the visual spectrum 
exceeded $\sim$ 50000 W.
Bolides are typically seen at altitudes $\sim 100\,$km. 
This implies a flux received at ground level of 
$F_{thresh} \geq 10^{-8}\,$W m$^{-2}$.

The DFN reported 
a limiting apparent magnitude  $m_V = 0.5$
\cite{Howie2017}. 
This is an instrument-specific threshold
for the faintest object that is reliably detectable.
We infer a minimum flux that must be received 
at the DFN camera of
$F_{thresh} \gtrsim 10^{-8}$W m$^{-2}$,
similar to the value derived above.

A meteor (including a macro transiting the atmosphere)
must exceed this threshold brightness 
to be detectable by the network cameras. 
This brightness depends on both 
the intrinsic luminosity of the meteor 
and its distance from the cameras.
We neglect the  effect of 
photon scattering along the path 
from the macro trajectory to the camera,
which is a small correction 
\cite{Sidhu:2018auv}.
The actual flux received is therefore
proportional to the 
intrinsic luminosity of the macro as it passes
through the atmosphere 
(which depends on $v_x$ and $\sigma_x$ for the macro), 
and inversely proportional to the square of the distance
 from the macro trajectory to the camera.
 As macros are expected to be significantly denser
than an ordinary meteorite, they are unlikely to
fragment at high altitude like ordinary meteorities.
Thus they are expected to be observable
much closer to the ground.

{\color{black}We treat the entire emission region
produced by the macro as a point source
within the field-of-view (FOV)
of a single camera pixel.
This is reasonable given that the pixels have a small FOV 
\cite{Howie2017} of $(0.036 \degree)^2$, i.e.
the transverse length of the path seen by a pixel
is significantly less than the distance between the macro and the
pixel.}

As a macro at altitude $D$
passes through the angle $\theta$ subtended 
by the FOV of a pixel, 
it traverses a distance $L=D\theta$ -- 
depositing energy  $ L dE/dx$ (per equation \eqref{dedx}),
over a time $L/v(x)$.
This creates a plasma, which persists for a time $t_{I0}$ \cite{Sidhu:2018auv}.
As the plasma cools it emits a fraction $\epsilon$ of its
energy into the part of the (visible) spectrum
to which the camera is sensitive.
The flux incident on the camera pixel is then 
\begin{widetext}
\begin{align}\label{thresh}
F &= \min\left(\frac{v(x)}{L},\frac{1}{t_{I0}}\right)
\epsilon \frac{dE}{dx} \frac{L}{4\pi D^2}  \nonumber
\\
&= 3 J m^{-2}~\min\left(\frac{v(x)}{L},\frac{1}{t_{I0}}\right)
 \left(\frac{\sigma_x}{cm^2}\right)^2
\left(\frac{v(x)}{250 kms^{-1}}\right)^4
e^{-\frac{3D}{20km}}\frac{km}{D}.
\end{align}
\end{widetext}
{\color{black}
For macros, we have defined an analogous quantity ${\cal L}$ as 
\begin{equation}
{\cal L} = \min\left(\frac{v(x)}{L},\frac{1}{t_{I0}}\right)
\epsilon \frac{dE}{dx} L
\end{equation}
where 
$\epsilon$ is the fraction of energy that emerges into visible wavelengths
previously calculated in \cite{1808.06978} 
for macros passing through the atmosphere, 
\begin{equation}\label{epsilon}
    \begin{aligned}
		\epsilon&=\frac{\mbox{N}_\gamma^{thin}\overline{E}}{\frac{1}{2}\rho v_x^2\sigma_{x}\mbox{L}}\\
		&\approx 2\times 10^2 \left(\frac{\sigma_x}{cm^2}\right)^2\left(\frac{v(x)}{250\,km\,s^{-1}}\right)^4 e^{-\frac{3D}{10\,km}}\,,
	\end{aligned}
\end{equation}
where $\mbox{N}_\gamma^{thin}$ is the number of photons emitted
by the plasma, and $\overline{E}$ is the average energy of those photons.}
We expect that at large cross-sections $\sigma_x \gtrapprox 2 \times 10^{-3}\,$
cm$^2$, $\epsilon$ would eventually saturate
at some fraction of the total energy deposited. 
(See section 2 of \cite{Sidhu:2018auv} 
for a discussion on why this is the case.)
We have divided
by the larger of the two timescales 
present in the problem, 
$\frac{L}{v_X}\,$, 
which is the pixel crossing time,
and $t_{I0}$, 
which is the time of existence of the plasma
produced by the macro. 
The larger of
the two timescales is what determines the flux
produced at the camera.  

Using 
\eqref{thresh} and \eqref{epsilon},
taking account of the distribution in altitude of macro trajectories,
and using the macro velocity distribution,
we can 
ascertain the detectability of macros 
of particular $\sigma_x$, 
by requiring $F \geq F_{thresh}$ sometime in
their passage through the atmosphere. 

To constrain $\sigma_x$ as a function of $M_x$, 
we enforce the null observation
of a fast-moving macro that
would have produced a detectable bolide. 
{\color{black}For macros large enough to be detected, 
the expected number of events 
that a survey should have seen is determined by
multiplying \eqref{fluxcondition} by the relevant
observation time
of the survey.}
We obtain $v_{thresh}$ from (\ref{thresh}) and the requirement that $F\geq F_{thresh}$,
taking care to note that $v_{thresh}$ 
is its value at the top of the atmosphere, 
not at altitude $D$.
Of course, $v_{thresh}$ depends {\color{black}on $D$ and so
we} must sample altitudes appropriately, 
as described below.

The passage of a macro through the field
of view of a survey
is a Poisson process. 
The probability $P(n)$ of 
$n$ passages over a given exposure time,
 follows the distribution:
\begin{equation}
\label{eq:Poisson}
P(n) = \frac{{N_{events}}^n}{n!}
e^{- N_{events}}\,.
\end{equation}
where $N_{events}$ is, as computed above, 
the expected
number of events per interval,
and varies by detector {\color{black}area} and exposure time.

If a macro were to cross the field
of view of one of these networks,
the network would observe a bolide moving 
far too fast 
to be bound within the Solar System.
This presents an easy way to
distinguish between bolides
formed from solar-system meteoroids
and macros.\footnote{{\color{black}Interstellar meteors
could provide false positives for macros. 
In 30 years of CNEOS data, it has been determined
that at most 1 such interstellar meteor
has been observed \cite{1904.07224},
with even this conclusion being uncertain
\cite{billings2019}. 
However, CNEOS is currently sensitive to objects 
that are $\gtrsim 140$m across, 
and so extrapolating abundances
to the much smaller objects we are considering is 
uncertain. 
We take the background rate to be zero,
but acknowledge that the detection of a fast-moving
bolide would require follow up investigation to
distinguish between a meteor and a macro.}}.
Thus, the non-observation of such
a fast-moving bolide
allows us to constrain macros that 
are big enough to have produced a
detectable signal.  Requiring
$N_{events} \geq 3$ (obtained
by setting the probability $P(n=0) \leq 0.05$
in (\ref{eq:Poisson})), 
gives us a 95\% C.L. contraint
that macros of at most the selected mass and 
at least the selected cross-section
do not constitute all the dark matter. 

\subsection*{Analysis}

To determine the 
regions of parameter space that are constrained or could 
be further probed we proceed as follows:
\begin{itemize}
\item Iterate $v_x$ over the range of allowed values ($0km/s$-$550km/s$).
\item For each value of $v_x$, 
iterate over a wide range of altitudes $10 m
\leq D \leq 100$km. 
The lower limit comes from the dependence of the event rate on $D$. 
Taking each camera to have a roughly conical field-of-view, 
and noting that the cameras are $\sim 100\,$km apart, 
$A_{det}$ will scale as $D^2$. 
For $D\leq 10\,$m, the event rate
is too low for any unconstrained parameter space to be probed. 

The upper bound comes from the strong dependence of
the incident flux $F$ on $D$, as seen in (\ref{thresh}).
For $D\gtrsim100\,$km, we find that the macro cross-section
necessary to produce a detectable signal is already 
ruled out by other considerations.

\item For each value of $D$, 
determine the smallest and largest 
cross-sections that can be probed 
from \eqref{thresh} (requiring that $F\geq F_{thresh}$),
and the maximum mass that may be probed
from \eqref{fluxcondition} (requiring that $N_{events}\geq 3$). 
The upper bound on the range of $\sigma$ that can be probed,
is always determined by by requiring macros
be sufficiently dense to survive passage through the atmosphere, which we conservatively take to be $10^3$ times atomic density.

\item From the above iterations, for each value of 
$M_x$ determine the range of values of $\sigma_x$ that has
been constrained or can  be probed.  

 These final values of $M_x$ and
$\sigma_x$ are presented  in
Figure 1.
\end{itemize}

\section{Constraints from past
meteorite networks}
We derive constraints on macros from a lack 
of visible evidence of them transiting
through the atmosphere across
the fields of view of two meteorite networks: 
a combination
of   the 
U.S. Prairie 
Network,
the Canadian Network, and the
Eastern European Network, 
which we refer to 
collectively as the PCE network,
which operated in the 60s, 70s and 80s;
\cite{Hills1986}
and the  the Desert Fireball Network,
\cite{Howie2017}
currently operating in Australia.

\subsection*{PCE network}

No fireballs moving fast enough to have 
an origin beyond the solar system were 
observed in a large-area survey over an 
``effective'' (i.e. scaled to the full 
area of the Earth) period of 30 hours 
\cite{Hills1986},
so $A_{\det} t \sim A_{\bigoplus} 30\,$hours.

Using \eqref{thresh} and \eqref{epsilon},
and requiring $F \geq F_{thresh}=10^{-8}\,$W m$^{-2}$, 
macro visibility by PCE network cameras requires
\begin{equation}\label{sigmaxdetectablebolide2}
\sigma_x \geq
2\times 10^{-4} {\mathrm {cm^2}} 
	\left(\frac{250 \mathrm {km/s}}{v_x}\right)^{2}
	\left(\frac{D}{\mathrm{km}}\right)^{1/2} \\e^{3D/20{\mathrm {km}}}.
\end{equation}
In general, 
the  plasma lifetime  $t_{I0}$ is larger than
the pixel crossing time of the macro, 
and is the term relevant in \eqref{thresh}.
Thus
\begin{equation} \label{Nevents}
		N_{events} 
		=1.9\times 10^{6} f_x f_{det}(\sigma_x)  \frac{g}{M_{x}}\,,
\end{equation}
where $f_x\equiv \Omega_x/\Omega_{DM}$ 
is the fraction of the dark matter in macros.
(We continue to assume that all macros have a
 single mass $M_x$ and cross-section $\sigma_x$.)

\begin{figure*}
 \centering
 \begin{minipage}[c]{\textwidth}
 \centering
        \includegraphics[width=7.0in]{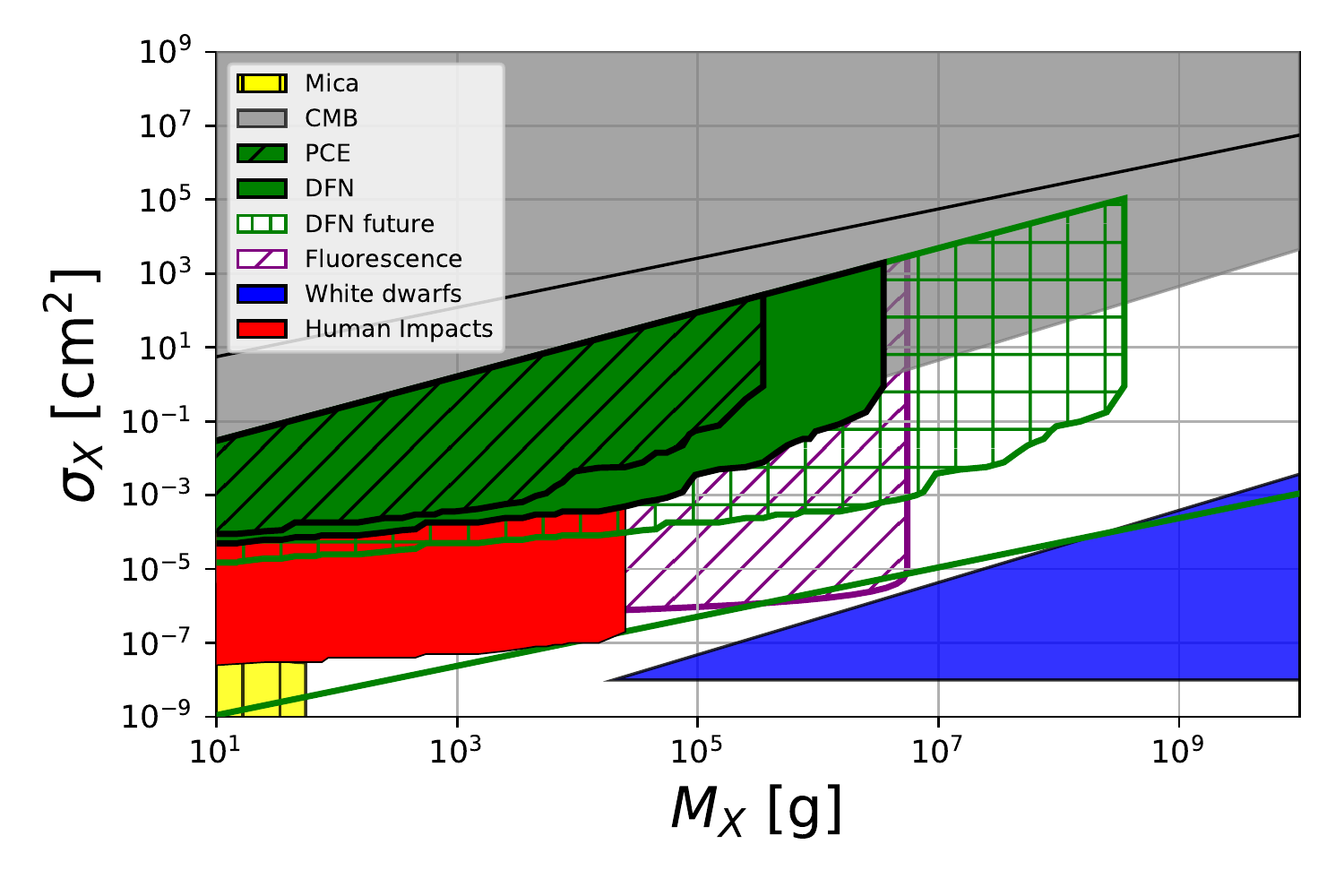}
        \caption{Constraints (solid green) derived
        from the null observation of
        bolides produced
        by a passing macro {\color{black}in the 
        combination of the
        U.S. Prairie, Canadian, and Eastern European (PCE)
        bolide networks that existed in the 1960's - 1980's, and the currently
        operational Desert Fireball Network (DFN),}
        and the region of parameter space
        that may eventually be probed 
        given certain reasonable assumptions
        about the final size of the DFN network 
        (green hatching). 
        The details of the green regions 
        are explained in the text.
        The region currently excluded by examination of ancient mica \cite{DeRujula:1984axn, Price:1988ge} is shown in yellow with vertical hatching;
		the grey region is excluded 
		from the effects of CMB photons 
		scattering off the macros \cite{1309.7588};
		the blue region from the continued existence of white dwarfs \cite{1805.07381};
		and the red from the lack of human impacts \cite{1907.06674}.  
		{\color{black}
		The purple (hatched) region is accessible in the future to the Fluorescence Detectors that are components of certain Ultra-High-Energy Cosmic Ray detectors e.g., those that may result
		from the JEM-EUSO program \cite{Sidhu:2018auv}.}}
        \label{fig:bolide}
 \end{minipage}
 \end{figure*}

The non-observation of any fast-moving fireballs, allows us to conclude, at the 95 per cent confidence level, that
\begin{equation}
f_x 
\leq \frac{M_{x}}{6 \times 10^{5} g} 
	\frac{1}{f_{det}(\sigma_x) }
\end{equation}
for macros satisfying \eqref{sigmaxdetectablebolide2}.
These results are presented
in Figure \ref{fig:bolide} in green
with black diagonal hatching.

\subsection*{Desert Fireball Network}
In this section we recalculate,
using the framework described above,
the region of parameter space
excluded by the non-observation
of an extra solar bolide by the
Desert Fireball Network (DFN).
DFN is an extensive array of cameras 
monitoring approximately one-third
of Australian skies for bolides
with a minimum magnitude
of $m_{\nu} \approx 0.5$
\cite{Howie2017}. 

The Australian Nullarbor plain is 
a good site for a fireball camera 
network
due to ideal viewing conditions.
The lack of vegetation and pale
geology make recovery of 
a fallen meteor easier.
These reasons,
along with the relative
ease of setting up
an extensive network
of cameras at a reasonable price
\cite{Howie2017}
led to creation
of the DFN over the last decade,
and its current state of
more than 50 cameras observing
most nights of the year 
\cite{fireballs}.
This large area will allow
new constraints
to be produced relative to the
previous network of cameras.

\subsubsection*{Current Constraints}

The expression for the minimum cross-
section that can be probed as
a function of $D$ and $v_x$
remains unchanged from \eqref{sigmaxdetectablebolide2},
as the specifics
of the cameras in the DFN and PCE
aren't significantly different,
as suggested by the equlaity of the two
threshold fluxes reported above.

The DFN array has been running with
an active detection area of 
$A_{det} \approx 2\times 10^6 {\mathrm km}^2$ for 
almost three years \cite{Howie2017}.
Thus the expected number of fireball events that the survey should have seen
is
\begin{equation} \label{DFN}
		N_{events} 
		=1.2\times 10^{7} f_x
		f_{det}(\sigma_x) \frac{g}{M_{x}}\,.
\end{equation}

We conclude that
\begin{equation}
f_x \leq \frac{M_{x}}{4 \times 10^{6} g}\frac{1}{f_{det}(\sigma_x)}
\end{equation}
for macros satisfying \eqref{sigmaxdetectablebolide2}.
These constraints are presented
in Figure \ref{fig:bolide} in green
with no hatching.

\subsubsection*{Future 
parameter space that may be probed}
There are concrete plans to form a 
global array of cameras to
increase the observing area for
meteors \cite{Howie2017}. 
The current array is observing
approximately $0.5\%$ of the
Earth's surface area. 

Since an important objective of 
such meteor-monitoring programs 
is to recover the remnant meteorite,
they are most likely to be extended over desert regions,
which also are the places most likely to provide
the highest fraction of times with
good viewing conditions. 
Deserts comprise approximately 10\% of the Earth's surface,
which is therefore an optimistic upper limit to the 
coverage of future arrays. 
Realistically, the planned duration of a 
meteor search is unlikely to be longer than 
30 years, a factor of 3 longer than the
current age of the DFN. 
We can thus expect at most a factor of 60 increase
in total exposure.

This yields 
\begin{equation} \label{DFNfuture}
		N_{events}
		=1.2\times 10^{9} f_x f_{det}(\sigma_x) \frac{g}{M_{x}}\,.
\end{equation}

The continued non-observation of fast-moving fireballs during an observing time 3 times
the current value
by an array 20 times
larger than
the current size would place
the constraint 
\begin{equation}
f_x \leq \frac{M_{x}}{4 \times 10^{8} g}\frac{1}{f_{det}(\sigma_x)}
\end{equation}
for macros satisfying \eqref{sigmaxdetectablebolide2}.
This region is presented in green
hatching in Figure \ref{fig:bolide}.
{\color{black}
In Figure \ref{fig:bolide}, 
we have also presented projections
from a recent study  \cite{Sidhu:2018auv} of the Fluorescence Detectors (FDs) that are components of large ultra-high-energy 
cosmic-ray detector arrays,
for comparison to the
results presented in this manuscript. While the
methods outlined here will allow us to probe higher
masses, the FDs will allow a greater sensitivity to
smaller cross-sections. Thus, a complimentary approach
will be ideal for probing as much of the parameter 
space as possible.}

The actual value of $M_x$ that will 
eventually be probed by the expanded DFN network
will of course depend
on the final size of the network
of cameras and its total live time. 
Values as high as 
$M_x \sim 10^{9}\,$g could be probed.

\section{Conclusion}
We have produced
constraints from the non-observation of ``extrasolar
meteors,'' i.e. bolides
produced by the passage of a macro
with a sufficiently large 
geometric cross
section $\sigma_x$. We have identified
a region of macro
parameter space
that could potentially be probed
by expansion
of the DFN network of meteorite 
cameras over the coming years.
This region 
represent a significant (up to $60$x)
improvement on the mass-reach of current constraints.

Since it is unclear what part
of the available parameter space macros should occupy,
it is vital to explore as much of it as possible.

The idea outlined here
is similar to the use of
Fluorescence Detectors
to look for photons produced
by a macro passing in the vicinity
\cite{1808.06978}. 
The minimum values of $\sigma_x$ that could
be probed by planned or potential expansions of the DFN
are not competitive with the lowest values of
$\sigma_x$ that could be probed using FDs
\cite{1808.06978}. 
However, it is very likely easier to reach higher masses
sooner with an upgraded bolide network than with an FD,
because of the relative (!) ease 
of expanding the existing  bolide networks \cite{Howie2017}
compared to building an appropriately configured FD network.
The FD network is needed to probe to lower $\sigma_x$,
especially to reach the potentially 
most interesting nuclear densities.

An interesting corollary to this
estimate of the future parameter space 
that could be probed by the DFN,
is that it seems unlikely 
that macro masses beyond 
$\sim 10^{9}\,$g could
be probed by any purpose-built
terrestrial detector 
assuming even an observation time of a century
and a target area the size of the Earth. 
Terrestrial probes 
(eg. ancient rocks \cite{DeRujula:1984axn,Price:1988ge,1905.10025}) 
could have been continuously exposed 
for up to $3\times10^9$ years,
but we are unlikely to carefully examine 
the more than 1km$^2$
that would be needed to push beyond $M_x=10^{9}$g.
It will therefore require innovative 
thinking about astrophysical probes
(eg. \cite{1805.07381})
to probe the very highest possible macro masses.

\bigskip
\acknowledgements
This work was partially supported by Department
of Energy grant DE-SC0009946 
to the particle astrophysics theory group at CWRU.

\bibliographystyle{apsrev4-1}
\bibliography{bolides}

\begin{thebibliography}{32}%
\makeatletter
\providecommand \@ifxundefined [1]{%
 \@ifx{#1\undefined}
}%
\providecommand \@ifnum [1]{%
 \ifnum #1\expandafter \@firstoftwo
 \else \expandafter \@secondoftwo
 \fi
}%
\providecommand \@ifx [1]{%
 \ifx #1\expandafter \@firstoftwo
 \else \expandafter \@secondoftwo
 \fi
}%
\providecommand \natexlab [1]{#1}%
\providecommand \enquote  [1]{``#1''}%
\providecommand \bibnamefont  [1]{#1}%
\providecommand \bibfnamefont [1]{#1}%
\providecommand \citenamefont [1]{#1}%
\providecommand \href@noop [0]{\@secondoftwo}%
\providecommand \href [0]{\begingroup \@sanitize@url \@href}%
\providecommand \@href[1]{\@@startlink{#1}\@@href}%
\providecommand \@@href[1]{\endgroup#1\@@endlink}%
\providecommand \@sanitize@url [0]{\catcode `\\12\catcode `\$12\catcode
  `\&12\catcode `\#12\catcode `\^12\catcode `\_12\catcode `\%12\relax}%
\providecommand \@@startlink[1]{}%
\providecommand \@@endlink[0]{}%
\providecommand \url  [0]{\begingroup\@sanitize@url \@url }%
\providecommand \@url [1]{\endgroup\@href {#1}{\urlprefix }}%
\providecommand \urlprefix  [0]{URL }%
\providecommand \Eprint [0]{\href }%
\providecommand \doibase [0]{http://dx.doi.org/}%
\providecommand \selectlanguage [0]{\@gobble}%
\providecommand \bibinfo  [0]{\@secondoftwo}%
\providecommand \bibfield  [0]{\@secondoftwo}%
\providecommand \translation [1]{[#1]}%
\providecommand \BibitemOpen [0]{}%
\providecommand \bibitemStop [0]{}%
\providecommand \bibitemNoStop [0]{.\EOS\space}%
\providecommand \EOS [0]{\spacefactor3000\relax}%
\providecommand \BibitemShut  [1]{\csname bibitem#1\endcsname}%
\let\auto@bib@innerbib\@empty
\bibitem [{\citenamefont {Witten}(1984)}]{PhysRevD.30.272}%
  \BibitemOpen
  \bibfield  {author} {\bibinfo {author} {\bibfnamefont {E.}~\bibnamefont
  {Witten}},\ }\href {\doibase 10.1103/physrevd.30.272} {\bibfield  {journal}
  {\bibinfo  {journal} {Physical Review D}\ }\textbf {\bibinfo {volume} {30}},\
  \bibinfo {pages} {272} (\bibinfo {year} {1984})}\BibitemShut {NoStop}%
\bibitem [{\citenamefont {Lynn}\ \emph {et~al.}(1990)\citenamefont {Lynn},
  \citenamefont {Nelson},\ and\ \citenamefont {Tetradis}}]{LYNN1990186}%
  \BibitemOpen
  \bibfield  {author} {\bibinfo {author} {\bibfnamefont {B.~W.}\ \bibnamefont
  {Lynn}}, \bibinfo {author} {\bibfnamefont {A.~E.}\ \bibnamefont {Nelson}}, \
  and\ \bibinfo {author} {\bibfnamefont {N.}~\bibnamefont {Tetradis}},\
  }\href@noop {} {\bibfield  {journal} {\bibinfo  {journal} {Nuclear Physics
  B}\ }\textbf {\bibinfo {volume} {345}},\ \bibinfo {pages} {186} (\bibinfo
  {year} {1990})}\BibitemShut {NoStop}%
\bibitem [{\citenamefont {Lynn}(2010)}]{1005.2124}%
  \BibitemOpen
  \bibfield  {author} {\bibinfo {author} {\bibfnamefont {B.~W.}\ \bibnamefont
  {Lynn}},\ }\href@noop {} {\enquote {\bibinfo {title} {Liquid phases in
  {SU(3)} {C}hiral {P}erturbation {T}heory: {D}rops of {S}trange {C}hiral
  {N}ucleon {L}iquid and {O}rdinary {C}hiral {H}eavy {N}uclear {L}iquid},}\ }
  (\bibinfo {year} {2010}),\ \Eprint {http://arxiv.org/abs/arXiv:1005.2124}
  {arXiv:1005.2124} \BibitemShut {NoStop}%
\bibitem [{\citenamefont {Nelson}(1990)}]{Nelson:1990iu}%
  \BibitemOpen
  \bibfield  {author} {\bibinfo {author} {\bibfnamefont {A.~E.}\ \bibnamefont
  {Nelson}},\ }\href {\doibase 10.1016/0370-2693(90)90429-A} {\bibfield
  {journal} {\bibinfo  {journal} {Physical Letters}\ }\textbf {\bibinfo
  {volume} {B240}},\ \bibinfo {pages} {179} (\bibinfo {year}
  {1990})}\BibitemShut {NoStop}%
\bibitem [{\citenamefont {Zhitnitsky}(2003)}]{Zhitnitsky2003}%
  \BibitemOpen
  \bibfield  {author} {\bibinfo {author} {\bibfnamefont {A.~R.}\ \bibnamefont
  {Zhitnitsky}},\ }\href {\doibase 10.1088/1475-7516/2003/10/010} {\bibfield
  {journal} {\bibinfo  {journal} {Journal of Cosmology and Astroparticle
  Physics}\ }\textbf {\bibinfo {volume} {2003}},\ \bibinfo {pages} {010}
  (\bibinfo {year} {2003})}\BibitemShut {NoStop}%
\bibitem [{\citenamefont {Jacobs}\ \emph {et~al.}(2014)\citenamefont {Jacobs},
  \citenamefont {Starkman},\ and\ \citenamefont {Lynn}}]{jacobs2015macro}%
  \BibitemOpen
  \bibfield  {author} {\bibinfo {author} {\bibfnamefont {D.~M.}\ \bibnamefont
  {Jacobs}}, \bibinfo {author} {\bibfnamefont {G.~D.}\ \bibnamefont
  {Starkman}}, \ and\ \bibinfo {author} {\bibfnamefont {B.~W.}\ \bibnamefont
  {Lynn}},\ }\href {\doibase 10.1093/mnras/stv774} {\  (\bibinfo {year}
  {2014}),\ 10.1093/mnras/stv774},\ \Eprint
  {http://arxiv.org/abs/arXiv:1410.2236} {arXiv:1410.2236} \BibitemShut
  {NoStop}%
\bibitem [{\citenamefont {Jacobs}\ \emph {et~al.}(2015)\citenamefont {Jacobs},
  \citenamefont {Weltman},\ and\ \citenamefont
  {Starkman}}]{jacobs2015resonant}%
  \BibitemOpen
  \bibfield  {author} {\bibinfo {author} {\bibfnamefont {D.~M.}\ \bibnamefont
  {Jacobs}}, \bibinfo {author} {\bibfnamefont {A.}~\bibnamefont {Weltman}}, \
  and\ \bibinfo {author} {\bibfnamefont {G.~D.}\ \bibnamefont {Starkman}},\
  }\href {\doibase 10.1103/physrevd.91.115023} {\bibfield  {journal} {\bibinfo
  {journal} {Physical Review D}\ }\textbf {\bibinfo {volume} {91}},\ \bibinfo
  {pages} {115023} (\bibinfo {year} {2015})}\BibitemShut {NoStop}%
\bibitem [{\citenamefont {Price}(1988)}]{Price:1988ge}%
  \BibitemOpen
  \bibfield  {author} {\bibinfo {author} {\bibfnamefont {P.~B.}\ \bibnamefont
  {Price}},\ }\href {\doibase 10.1103/PhysRevD.38.3813} {\bibfield  {journal}
  {\bibinfo  {journal} {Physical Review D}\ }\textbf {\bibinfo {volume} {38}},\
  \bibinfo {pages} {3813} (\bibinfo {year} {1988})}\BibitemShut {NoStop}%
\bibitem [{\citenamefont {De~Rujula}\ and\ \citenamefont
  {Glashow}(1984)}]{DeRujula:1984axn}%
  \BibitemOpen
  \bibfield  {author} {\bibinfo {author} {\bibfnamefont {A.}~\bibnamefont
  {De~Rujula}}\ and\ \bibinfo {author} {\bibfnamefont {S.~L.}\ \bibnamefont
  {Glashow}},\ }\href {\doibase 10.1038/312734a0} {\bibfield  {journal}
  {\bibinfo  {journal} {Nature}\ }\textbf {\bibinfo {volume} {312}},\ \bibinfo
  {pages} {734} (\bibinfo {year} {1984})}\BibitemShut {NoStop}%
\bibitem [{\citenamefont {Alcock}\ \emph {et~al.}(2001)\citenamefont {Alcock}
  \emph {et~al.}}]{Alcock2001}%
  \BibitemOpen
  \bibfield  {author} {\bibinfo {author} {\bibfnamefont {C.}~\bibnamefont
  {Alcock}} \emph {et~al.},\ }\href {\doibase 10.1086/319636} {\bibfield
  {journal} {\bibinfo  {journal} {The Astrophysical Journal}\ }\textbf
  {\bibinfo {volume} {550}},\ \bibinfo {pages} {L169} (\bibinfo {year}
  {2001})}\BibitemShut {NoStop}%
\bibitem [{\citenamefont {Griest}\ \emph {et~al.}(2013)\citenamefont {Griest},
  \citenamefont {Cieplak},\ and\ \citenamefont {Lehner}}]{Griest2013}%
  \BibitemOpen
  \bibfield  {author} {\bibinfo {author} {\bibfnamefont {K.}~\bibnamefont
  {Griest}}, \bibinfo {author} {\bibfnamefont {A.~M.}\ \bibnamefont {Cieplak}},
  \ and\ \bibinfo {author} {\bibfnamefont {M.~J.}\ \bibnamefont {Lehner}},\
  }\href {\doibase 10.1103/physrevlett.111.181302} {\bibfield  {journal}
  {\bibinfo  {journal} {Physical Review Letters}\ }\textbf {\bibinfo {volume}
  {111}},\ \bibinfo {pages} {181302} (\bibinfo {year} {2013})}\BibitemShut
  {NoStop}%
\bibitem [{\citenamefont {Tisserand}\ \emph {et~al.}(2007)\citenamefont
  {Tisserand} \emph {et~al.}}]{astro-ph/0607207}%
  \BibitemOpen
  \bibfield  {author} {\bibinfo {author} {\bibfnamefont {P.}~\bibnamefont
  {Tisserand}} \emph {et~al.},\ }\href {\doibase 10.1051/0004-6361:20066017}
  {\bibfield  {journal} {\bibinfo  {journal} {Astronomy {\&} Astrophysics}\
  }\textbf {\bibinfo {volume} {469}},\ \bibinfo {pages} {387} (\bibinfo {year}
  {2007})}\BibitemShut {NoStop}%
\bibitem [{\citenamefont {Carr}\ \emph {et~al.}(2010)\citenamefont {Carr},
  \citenamefont {Kohri}, \citenamefont {Sendouda},\ and\ \citenamefont
  {Yokoyama}}]{0912.5297}%
  \BibitemOpen
  \bibfield  {author} {\bibinfo {author} {\bibfnamefont {B.~J.}\ \bibnamefont
  {Carr}}, \bibinfo {author} {\bibfnamefont {K.}~\bibnamefont {Kohri}},
  \bibinfo {author} {\bibfnamefont {Y.}~\bibnamefont {Sendouda}}, \ and\
  \bibinfo {author} {\bibfnamefont {J.}~\bibnamefont {Yokoyama}},\ }\href
  {\doibase 10.1103/physrevd.81.104019} {\bibfield  {journal} {\bibinfo
  {journal} {Physical Review D}\ }\textbf {\bibinfo {volume} {81}},\ \bibinfo
  {pages} {104019} (\bibinfo {year} {2010})}\BibitemShut {NoStop}%
\bibitem [{\citenamefont {Wilkinson}\ \emph {et~al.}(2014)\citenamefont
  {Wilkinson}, \citenamefont {Lesgourgues},\ and\ \citenamefont
  {B{\oe}hm}}]{1309.7588}%
  \BibitemOpen
  \bibfield  {author} {\bibinfo {author} {\bibfnamefont {R.~J.}\ \bibnamefont
  {Wilkinson}}, \bibinfo {author} {\bibfnamefont {J.}~\bibnamefont
  {Lesgourgues}}, \ and\ \bibinfo {author} {\bibfnamefont {C.}~\bibnamefont
  {B{\oe}hm}},\ }\href {\doibase 10.1088/1475-7516/2014/04/026} {\bibfield
  {journal} {\bibinfo  {journal} {Journal of Cosmology and Astroparticle
  Physics}\ }\textbf {\bibinfo {volume} {2014}},\ \bibinfo {pages} {026}
  (\bibinfo {year} {2014})}\BibitemShut {NoStop}%
\bibitem [{\citenamefont {B{\oe}hm}\ \emph {et~al.}(2001)\citenamefont
  {B{\oe}hm}, \citenamefont {Fayet},\ and\ \citenamefont
  {Schaeffer}}]{Bhm2001}%
  \BibitemOpen
  \bibfield  {author} {\bibinfo {author} {\bibfnamefont {C.}~\bibnamefont
  {B{\oe}hm}}, \bibinfo {author} {\bibfnamefont {P.}~\bibnamefont {Fayet}}, \
  and\ \bibinfo {author} {\bibfnamefont {R.}~\bibnamefont {Schaeffer}},\ }\href
  {\doibase 10.1016/s0370-2693(01)01060-7} {\bibfield  {journal} {\bibinfo
  {journal} {Physics Letters B}\ }\textbf {\bibinfo {volume} {518}},\ \bibinfo
  {pages} {8} (\bibinfo {year} {2001})}\BibitemShut {NoStop}%
\bibitem [{\citenamefont {Graham}\ \emph {et~al.}(2018)\citenamefont {Graham},
  \citenamefont {Janish}, \citenamefont {Narayan}, \citenamefont {Rajendran},\
  and\ \citenamefont {Riggins}}]{1805.07381}%
  \BibitemOpen
  \bibfield  {author} {\bibinfo {author} {\bibfnamefont {P.~W.}\ \bibnamefont
  {Graham}}, \bibinfo {author} {\bibfnamefont {R.}~\bibnamefont {Janish}},
  \bibinfo {author} {\bibfnamefont {V.}~\bibnamefont {Narayan}}, \bibinfo
  {author} {\bibfnamefont {S.}~\bibnamefont {Rajendran}}, \ and\ \bibinfo
  {author} {\bibfnamefont {P.}~\bibnamefont {Riggins}},\ }\href {\doibase
  10.1103/physrevd.98.115027} {\bibfield  {journal} {\bibinfo  {journal}
  {Physical Review D}\ }\textbf {\bibinfo {volume} {98}},\ \bibinfo {pages}
  {115027} (\bibinfo {year} {2018})}\BibitemShut {NoStop}%
\bibitem [{\citenamefont {Sidhu}\ \emph
  {et~al.}(2019{\natexlab{a}})\citenamefont {Sidhu}, \citenamefont {Scherrer},\
  and\ \citenamefont {Starkman}}]{1907.06674}%
  \BibitemOpen
  \bibfield  {author} {\bibinfo {author} {\bibfnamefont {J.~S.}\ \bibnamefont
  {Sidhu}}, \bibinfo {author} {\bibfnamefont {R.~J.}\ \bibnamefont {Scherrer}},
  \ and\ \bibinfo {author} {\bibfnamefont {G.}~\bibnamefont {Starkman}},\
  }\href@noop {} {\enquote {\bibinfo {title} {Death by dark matter},}\ }
  (\bibinfo {year} {2019}{\natexlab{a}}),\ \Eprint
  {http://arxiv.org/abs/arXiv:1907.06674} {arXiv:1907.06674} \BibitemShut
  {NoStop}%
\bibitem [{\citenamefont {Sidhu}\ \emph {et~al.}(2018)\citenamefont {Sidhu},
  \citenamefont {Abraham}, \citenamefont {Covault},\ and\ \citenamefont
  {Starkman}}]{1808.06978}%
  \BibitemOpen
  \bibfield  {author} {\bibinfo {author} {\bibfnamefont {J.~S.}\ \bibnamefont
  {Sidhu}}, \bibinfo {author} {\bibfnamefont {R.~M.}\ \bibnamefont {Abraham}},
  \bibinfo {author} {\bibfnamefont {C.}~\bibnamefont {Covault}}, \ and\
  \bibinfo {author} {\bibfnamefont {G.}~\bibnamefont {Starkman}},\ }\href
  {\doibase 10.1088/1475-7516/2019/02/037} {\  (\bibinfo {year} {2018}),\
  10.1088/1475-7516/2019/02/037},\ \Eprint
  {http://arxiv.org/abs/arXiv:1808.06978} {arXiv:1808.06978} \BibitemShut
  {NoStop}%
\bibitem [{\citenamefont {Abraham}\ \emph {et~al.}(2010)\citenamefont {Abraham}
  \emph {et~al.}}]{PAOFD}%
  \BibitemOpen
  \bibfield  {author} {\bibinfo {author} {\bibfnamefont {J.}~\bibnamefont
  {Abraham}} \emph {et~al.},\ }\href {\doibase 10.1016/j.nima.2010.04.023}
  {\bibfield  {journal} {\bibinfo  {journal} {Nuclear Instruments and Methods
  in Physics Research Section A: Accelerators, Spectrometers, Detectors and
  Associated Equipment}\ }\textbf {\bibinfo {volume} {620}},\ \bibinfo {pages}
  {227} (\bibinfo {year} {2010})}\BibitemShut {NoStop}%
\bibitem [{\citenamefont {Sidhu}\ \emph
  {et~al.}(2019{\natexlab{b}})\citenamefont {Sidhu}, \citenamefont {Starkman},\
  and\ \citenamefont {Harvey}}]{1905.10025}%
  \BibitemOpen
  \bibfield  {author} {\bibinfo {author} {\bibfnamefont {J.~S.}\ \bibnamefont
  {Sidhu}}, \bibinfo {author} {\bibfnamefont {G.}~\bibnamefont {Starkman}}, \
  and\ \bibinfo {author} {\bibfnamefont {R.}~\bibnamefont {Harvey}},\ }\href
  {\doibase 10.1103/physrevd.100.103015} {\bibfield  {journal} {\bibinfo
  {journal} {Physical Review D}\ }\textbf {\bibinfo {volume} {100}} (\bibinfo
  {year} {2019}{\natexlab{b}}),\ 10.1103/physrevd.100.103015}\BibitemShut
  {NoStop}%
\bibitem [{\citenamefont {Hills}(1986)}]{Hills1986}%
  \BibitemOpen
  \bibfield  {author} {\bibinfo {author} {\bibfnamefont {J.~G.}\ \bibnamefont
  {Hills}},\ }\href {\doibase 10.1086/114189} {\bibfield  {journal} {\bibinfo
  {journal} {The Astronomical Journal}\ }\textbf {\bibinfo {volume} {92}},\
  \bibinfo {pages} {595} (\bibinfo {year} {1986})}\BibitemShut {NoStop}%
\bibitem [{\citenamefont {Howie}\ \emph {et~al.}(2017)\citenamefont {Howie},
  \citenamefont {Paxman}, \citenamefont {Bland}, \citenamefont {Towner},
  \citenamefont {Cupak}, \citenamefont {Sansom},\ and\ \citenamefont
  {Devillepoix}}]{Howie2017}%
  \BibitemOpen
  \bibfield  {author} {\bibinfo {author} {\bibfnamefont {R.~M.}\ \bibnamefont
  {Howie}}, \bibinfo {author} {\bibfnamefont {J.}~\bibnamefont {Paxman}},
  \bibinfo {author} {\bibfnamefont {P.~A.}\ \bibnamefont {Bland}}, \bibinfo
  {author} {\bibfnamefont {M.~C.}\ \bibnamefont {Towner}}, \bibinfo {author}
  {\bibfnamefont {M.}~\bibnamefont {Cupak}}, \bibinfo {author} {\bibfnamefont
  {E.~K.}\ \bibnamefont {Sansom}}, \ and\ \bibinfo {author} {\bibfnamefont
  {H.~A.~R.}\ \bibnamefont {Devillepoix}},\ }\href {\doibase
  10.1007/s10686-017-9532-7} {\bibfield  {journal} {\bibinfo  {journal}
  {Experimental Astronomy}\ }\textbf {\bibinfo {volume} {43}},\ \bibinfo
  {pages} {237} (\bibinfo {year} {2017})}\BibitemShut {NoStop}%
\bibitem [{Note1()}]{Note1}%
  \BibitemOpen
  \bibinfo {note} {This is the distribution of macro velocities in a
  non-orbiting frame moving with the Galaxy. When considering the velocity of
  macros impacting the atmosphere, \protect \textup {\hbox {\mathsurround \z@
  \protect \normalfont (\ignorespaces \ref {eq:maxwellian}\unskip \@@italiccorr
  )}} is modified by the motion of the Sun and Earth in that frame, and by the
  Sun's and Earth's gravitational potential. We have taken into account these
  effects (as explained, for example, in \cite {Freese2013}), except the
  negligible effect of Earth's gravitational potential.}\BibitemShut {Stop}%
\bibitem [{\citenamefont {Bovy}\ and\ \citenamefont
  {Tremaine}(2012)}]{Bovy2012}%
  \BibitemOpen
  \bibfield  {author} {\bibinfo {author} {\bibfnamefont {J.}~\bibnamefont
  {Bovy}}\ and\ \bibinfo {author} {\bibfnamefont {S.}~\bibnamefont
  {Tremaine}},\ }\href {\doibase 10.1088/0004-637x/756/1/89} {\bibfield
  {journal} {\bibinfo  {journal} {The Astrophysical Journal}\ }\textbf
  {\bibinfo {volume} {756}},\ \bibinfo {pages} {89} (\bibinfo {year}
  {2012})}\BibitemShut {NoStop}%
\bibitem [{\citenamefont {Mccrosky}\ and\ \citenamefont
  {Boeschenstein}(1965)}]{mccrosky_boeschenstein_1965}%
  \BibitemOpen
  \bibfield  {author} {\bibinfo {author} {\bibfnamefont {R.~E.}\ \bibnamefont
  {Mccrosky}}\ and\ \bibinfo {author} {\bibfnamefont {J.~H.}\ \bibnamefont
  {Boeschenstein}},\ }\href {\doibase 10.1117/12.7971304} {\bibfield  {journal}
  {\bibinfo  {journal} {Optical Engineering}\ }\textbf {\bibinfo {volume} {3}}
  (\bibinfo {year} {1965}),\ 10.1117/12.7971304}\BibitemShut {NoStop}%
\bibitem [{\citenamefont {Opik}(1959)}]{1959}%
  \BibitemOpen
  \bibfield  {author} {\bibinfo {author} {\bibfnamefont {E.~J.}\ \bibnamefont
  {Opik}},\ }\href {\doibase 10.1002/qj.49708536526} {\bibfield  {journal}
  {\bibinfo  {journal} {Quarterly Journal of the Royal Meteorological Society}\
  }\textbf {\bibinfo {volume} {85}},\ \bibinfo {pages} {320} (\bibinfo {year}
  {1959})}\BibitemShut {NoStop}%
\bibitem [{\citenamefont {Sidhu}\ \emph
  {et~al.}(2019{\natexlab{c}})\citenamefont {Sidhu}, \citenamefont {Abraham},
  \citenamefont {Covault},\ and\ \citenamefont {Starkman}}]{Sidhu:2018auv}%
  \BibitemOpen
  \bibfield  {author} {\bibinfo {author} {\bibfnamefont {J.~S.}\ \bibnamefont
  {Sidhu}}, \bibinfo {author} {\bibfnamefont {R.~M.}\ \bibnamefont {Abraham}},
  \bibinfo {author} {\bibfnamefont {C.}~\bibnamefont {Covault}}, \ and\
  \bibinfo {author} {\bibfnamefont {G.}~\bibnamefont {Starkman}},\ }\href
  {\doibase 10.1088/1475-7516/2019/02/037} {\bibfield  {journal} {\bibinfo
  {journal} {JCAP}\ }\textbf {\bibinfo {volume} {1902}},\ \bibinfo {pages}
  {037} (\bibinfo {year} {2019}{\natexlab{c}})},\ \Eprint
  {http://arxiv.org/abs/1808.06978} {arXiv:1808.06978 [astro-ph.HE]}
  \BibitemShut {NoStop}%
\bibitem [{Note2()}]{Note2}%
  \BibitemOpen
  \bibinfo {note} {{\protect \color {black}Interstellar meteors could provide
  false positives for macros. In 30 years of CNEOS data, it has been determined
  that at most 1 such interstellar meteor has been observed \cite {1904.07224},
  with even this conclusion being uncertain \cite {billings2019}. However,
  CNEOS is currently sensitive to objects that are $\gtrsim 140$m across, and
  so extrapolating abundances to the much smaller objects we are considering is
  uncertain. We take the background rate to be zero, but acknowledge that the
  detection of a fast-moving bolide would require follow up investigation to
  distinguish between a meteor and a macro.}}\BibitemShut {Stop}%
\bibitem [{fir()}]{fireballs}%
  \BibitemOpen
  \href@noop {} {\enquote {\bibinfo {title} {The research - fireballs in the
  sky},}\ }\bibinfo {howpublished}
  {\url{http://fireballsinthesky.com.au/the-research/}},\ \bibinfo {note}
  {accessed: 2019-06-13}\BibitemShut {NoStop}%
\bibitem [{\citenamefont {Freese}\ \emph {et~al.}(2013)\citenamefont {Freese},
  \citenamefont {Lisanti},\ and\ \citenamefont {Savage}}]{Freese2013}%
  \BibitemOpen
  \bibfield  {author} {\bibinfo {author} {\bibfnamefont {K.}~\bibnamefont
  {Freese}}, \bibinfo {author} {\bibfnamefont {M.}~\bibnamefont {Lisanti}}, \
  and\ \bibinfo {author} {\bibfnamefont {C.}~\bibnamefont {Savage}},\ }\href
  {\doibase 10.1103/revmodphys.85.1561} {\bibfield  {journal} {\bibinfo
  {journal} {Reviews of Modern Physics}\ }\textbf {\bibinfo {volume} {85}},\
  \bibinfo {pages} {1561} (\bibinfo {year} {2013})}\BibitemShut {NoStop}%
\bibitem [{\citenamefont {Siraj}\ and\ \citenamefont
  {Loeb}(2019)}]{1904.07224}%
  \BibitemOpen
  \bibfield  {author} {\bibinfo {author} {\bibfnamefont {A.}~\bibnamefont
  {Siraj}}\ and\ \bibinfo {author} {\bibfnamefont {A.}~\bibnamefont {Loeb}},\
  }\href@noop {} {\enquote {\bibinfo {title} {Discovery of a meteor of
  interstellar origin},}\ } (\bibinfo {year} {2019}),\ \Eprint
  {http://arxiv.org/abs/arXiv:1904.07224} {arXiv:1904.07224} \BibitemShut
  {NoStop}%
\bibitem [{\citenamefont {Billings}(2019)}]{billings2019}%
  \BibitemOpen
  \bibfield  {author} {\bibinfo {author} {\bibfnamefont {L.}~\bibnamefont
  {Billings}},\ }\href@noop {} {\enquote {\bibinfo {title} {Did a meteor from
  another star strike earth in 2014?}}\ }\bibinfo {howpublished}
  {https://www.scientificamerican.com/article/did-a-meteor-from-another-star-strike-earth-in-2014/}
  (\bibinfo {year} {2019}),\ \bibinfo {note} {{A}ccessed: October 23rd
  2019}\BibitemShut {NoStop}%
\end{thebibliography}%

\end{document}